\documentclass[conference,a4paper]{IEEEtran}
\usepackage{xcolor}
\usepackage{balance}
\usepackage{amsmath,flushend}
\ifCLASSINFOpdf
  \usepackage[pdftex]{graphicx}
\else
  \usepackage[dvips]{graphicx}
\fi
\ifCLASSOPTIONcompsoc
 \usepackage[caption=false,font=normalsize,labelfont=sf,textfont=sf]{subfig}
\else
 \usepackage[caption=false,font=footnotesize]{subfig}
\fi

\begin{document}
\title{Reconfigurable Intelligent Surfaces for THz: \\Hardware Design and Signal Processing Challenges}
\author{\IEEEauthorblockN{
George C. Alexandropoulos$^1$, Antonio Clemente$^2$,  
S\'{e}rgio Matos$^3$, Ryan Husbands$^4$, \\ Sean Ahearne$^5$, Qi Luo$^6$, Verónica Lain-Rubio$^7$, Thomas K\"{u}rner$^8$, and Lu\'{i}s M. Pessoa$^9$
}                   
\IEEEauthorblockA{$^1$National and Kapodistrian University of Athens, Greece, $^2$CEA-Leti, France,$^3$University Institute of Lisbon, Portugal,\\
$^4$British Telecom, UK, $^5$Dell Technologies, Ireland, $^6$University of Hertfordshire, UK,\\ $^7$ACST, Germany, $^8$Braunschweig Technical University, Germany, $^9$INESC TEC, Portugal, }
\IEEEauthorblockA{e-mails: alexandg@di.uoa.gr, luis.m.pessoa@inesctec.pt}
}

\maketitle

\begin{abstract}
Wireless communications in the THz frequency band is an envisioned revolutionary technology for sixth Generation (6G) networks. However, such frequencies impose certain coverage and device design challenges that need to be efficiently overcome. To this end, the development of cost- and energy-efficient approaches for scaling these networks to realistic scenarios constitute a necessity. Among the recent research trends contributing to these objectives belongs the technology of Reconfigurable Intelligent Surfaces (RISs). In fact, several high-level descriptions of THz systems based on RISs have been populating the literature. Nevertheless, hardware implementations of those systems are still very scarce, and not at the scale intended for most envisioned THz scenarios. In this paper, we overview some of the most significant hardware design and signal processing challenges with THz RISs, and present a preliminary analysis of their impact on the overall link budget and system performance, conducted in the framework of the ongoing TERRAMETA project.
\end{abstract}

\vskip0.5\baselineskip

\section{Introduction}
The concept of Reconfigurable Intelligent Surfaces (RISs) has been recently established as a new wireless connectivity paradigm \cite{Strinati_2021a} for the upcoming sixth Generation (6G) of broadband networks, envisioning dynamic control of the radio propagation environment~\cite{alexandg_2021} to improve, and even enable with reduced/efficient hardware infrastructure~\cite{Tsinghua_RIS_Tutorial,RISoverview2023}, a wide variety of wireless communications~\cite{Alexandropoulos2022Pervasive}, localization~\cite{Keykhosravi2022infeasible}, and integrated sensing and communications applications~\cite{RIS_ISAC_SPM}. While the readiness of the technology is still ongoing the transition from theoretical works and new ideas to precise definitions and boundaries~\cite{EURASIP_RIS,ETSI}, the key feature that characterizes an RIS is its dynamic reflective response that can be realized with nearly passive hardware components~\cite{Tsinghua_RIS_Tutorial}, thus, exhibiting the potential to be very efficient in terms of energy consumption~\cite{huang2019reconfigurable}. However, the actual benefits of RISs, in terms of cost and overall efficiency, require rigorous quantification, potentially depend on the actual application and the operating frequency, and are tightly coupled to other alternative existing technologies, such as multi-antenna relays with hybrid analog and digital beamforming~\cite{hybrid_relays} and phased antenna arrays~\cite{bjornson2020reconfigurable}. 

Another candidate technology for 6G, which targets extensive unlicensed bandwidth to support rate- and sensing-demanding immersive applications with increased confidentiality, robustness to interference, and reduced latency, is terahertz (THz) communications (in the range of $0.1-10$ THz)~\cite{6gvision,ETSI_THz}. However, similar to the RIS technology, this new frequency band is accompanied with two key challenges when considered for wireless operations. The first is the high penetration loss, which can be critical even for very small link distances, and the second concerns the design of cost- and energy-efficient compact transceivers. The current trend to counteract the reduced coverage of THz communications leverages extremely large Multiple-Input Multiple-Output (MIMO) systems~\cite{MRV17_MIMO_Survey,Shlezinger2021Dynamic,HMIMO_survey}, which are capable of realizing highly directive beamforming. It needs to be noted, however, that due to the ultra-high bandwidth available for THz communications, the propagation delay across the large antenna arrays at the communication terminals can exceed the sampling period. In addition, as the antenna array size increases, its near-field range gets expanded, hence, the phase of the array steering vector becomes nonlinear to the antenna index, which gives birth to the beam-squint effect~\cite{8882325,Vlachos_beamsquint_2023}.

In this paper, we overview the research framework and objectives of the ongoing TERRAMETA project~\cite{TERRAMETA_website}, which capitalizes on the RIS-empowered wireless connectivity paradigm to realize efficient outdoor and indoor THz communication systems, targeting to devise novel signal processing techniques and hardware designs for THz RISs. A preliminary analysis of the impact of RISs on the overall THz link budget and system performance is presented, together with challenges for THz RIS hardware design and beamforming as well as signal processing with multi-functional RISs. 

The paper is organized as follows: Section~II describes relevant outdoor and indoor THz communications use cases that can be impacted by RIS, and includes a link budget analytical formulation enabling the determination of the required RIS aperture size for a given scenario. Section~III describes technological challenges associated with THz RIS implementation, and finally, Section IV provides the conclusions.


\section{RIS-Enabled THz Use Cases}\label{sec:THz_use_cases}
\begin{figure}[t!]
	\centering
	\includegraphics[width=0.9\linewidth]{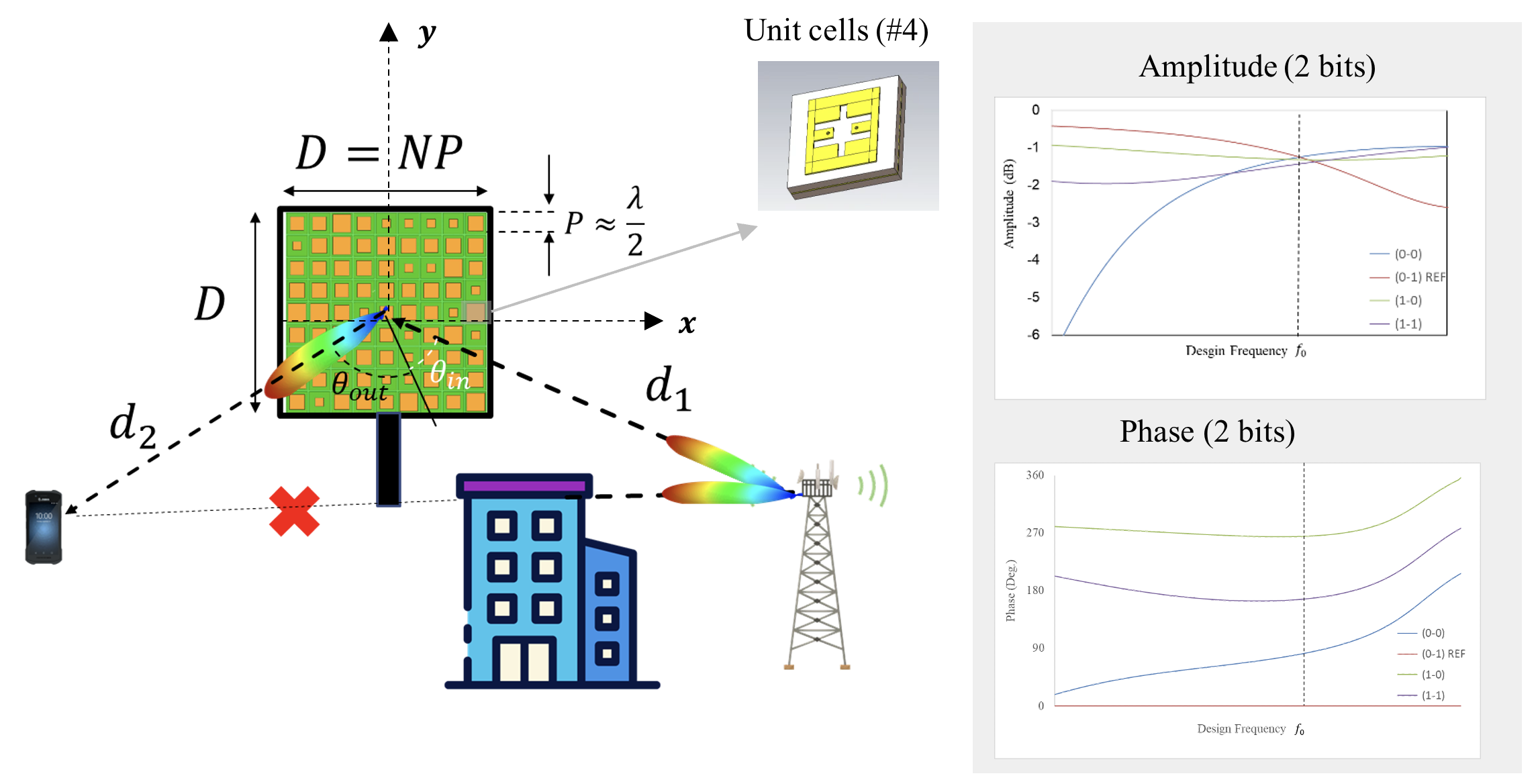}
	\caption{\small{Non-light of sight communications enabled by an RIS. An example response of a realistic RIS unit cell with $2$-bit amplitude and phase quantization is also illustrated.}}
	\label{fig:NLOS_RIS}
\end{figure}
Non-line-of-sight (LOS) communications enabled by an RIS is one of the most popular applications for outdoor scenarios~\cite{EURASIP_RIS}. The idea is that the surface dynamically redirects (by means of anomalous reflection) the incoming wave of the Base Station (BS) to the mobile terminal (and vice-versa), as depicted in Fig.~\ref{fig:NLOS_RIS}. 

\subsection{Outdoor Use Case: Enhancing Coverage to Black Spots}
For mobile network operators, ensuring ubiquitous and seamless coverage is paramount, especially when transitioning to higher frequency bands, such as THz, that inherently have higher propagation losses~\cite{9794668}. One significant challenge for operators is the existence of coverage black spots. Traditional methods involve deploying additional BSs to bridge these gaps, however, this solution is not only expensive, but may also result, in increased energy consumption, infrastructure management costs, and other negative effects. RISs, with their dynamic reflection capabilities, offer a cost-effective and power-efficient solution to this problem. In a scenario, such as a dense urban environment, where LOS communication from a BS is impossible (see Fig.~\ref{fig:NLOS_RIS}), instead of setting up an additional BS, an RIS could be strategically placed on the side of a building or a light pole. This RIS would then dynamically redirect the incoming waves from the BS, navigating them around buildings and other obstacles and towards the receiving device (and vice-versa), thus, providing coverage to an area that was previously a black spot. This approach, not only reduces the deployment and operational costs, but also minimizes potential interference by using directional beams.

\subsection{Indoor Use Case: Ensuring High-Speed Connectivity for Industry 4.0 and Beyond}
As we transition towards Industry 4.0 and beyond, factories are gradually becoming more software-defined with an increasing reliance on mobile robots and drones equipped with high-resolution sensors ~\cite{nayak2015software}. Those robots and drones are expected to generate massive amounts of wireless data, and therefore, the demand for dense large-bandwidth communications across the factory floor is increasing. THz communications, with the support of RISs, can be pivotal in enabling such dense communications environments. Consider a large factory floor, where numerous robots and drones are operating simultaneously generating potentially hundreds of Gigabits per second of sensor data that need to be processed, in many cases in real time. Traditional Wi-Fi, or even 5G solutions, might not suffice due to the sheer volume of data transfer and the need for ultra-low latency between devices. Similar to an outdoor scenario, it is not practical to deploy a THz BS to cover every black spot behind every factory machine that exists on the floor. In this use case, one or more THz RISs can be strategically placed the factory walls and ceiling to ensure that THz waves are optimally reflected towards the necessary nodes, ensuring seamless and high-speed connectivity. 

\subsection{An RIS Design Example at THz}\label{sec:link_budget}
The structure and layout of an RIS needs to be carefully dedigned in order to enable the previously discussed THz use cases. Its aperture is composed by unit cells that act as filters that control locally the phase of the outgoing waves~\cite{alexandg_2021,Tsinghua_RIS_Tutorial}. Each unit cell has a set of discrete states (the inset of Fig.~\ref{fig:NLOS_RIS} shows an example with $4$ states, i.e., $2$ bits) that interchange through a biasing network that actively controls the switching mechanism. The number of switches will impact on the power consumption and complexity/cost of the metasurface. Therefore, typical RIS hardware designs are based on unit cells with few bits ($1$ or $2$ bits) that can provide $360^{\rm o}$ phase coverage for a given design frequency $f_0$. For a proper analysis, it is important to get a realistic estimate of the targeted size of the RIS required to meet the link budget. In the following, we provide a simple way of retrieving this value while considering typical values for sub-THz communications. We consider a distance link of $100$ meters at $140$ GHz with an RIS placed at equal distance between a receiving terminal and the BS ($d_1=d_2=50$ meters in Fig.~\ref{fig:NLOS_RIS}). This problem can be formulated using classical bi-static analysis, yielding the following receive/transmit power ratio:
\begin{equation}\label{eq:ratio}
\begin{aligned}
    \left( \frac{P_{\rm rx}}{P_{\rm tx}} \right)_{\rm dB} =\, & G_{\rm BS}^{\rm dB} + G_{\rm T}^{\rm dB} + (\sigma_{\rm RIS})_{\rm dB} \\
    &+ 10\log_{10}\left( \frac{1}{(4\pi)^3}\left( \frac{\lambda}{d_1 d_2} \right)^2 \right),
\end{aligned}    
\end{equation}
where the antenna gain and transmit power of the BS are $G_{\rm BS}^{\rm dB}=46$dBi and $P_{\rm tx}=20$ dBm, respectively, the RIS incident and output angles are considered as $\theta_{\rm in}=0^{\rm o}$ and $\theta_{\rm out}=45^{\rm o}$, respectively, and the antenna gain of use terminal is assumed as $G_{\rm T}^{\rm dB}=10$ dBi with the minimum received power (terminal sensitivity) being $P_{\rm rx}=-60$ dBm (considering $4$-ary Quadrature Amplitude Modulation (QAM), $7$ dB of noise figure, and $2$ GHz of signal bandwidth). The unknown variable of the desired RIS in \eqref{eq:ratio} is its Radar Cross Section (RCS), $\sigma_{\rm RIS}$, which is bounded by the specular reflection limit in a Perfect Electric Conductor (PEC) surface, and can be estimated as a function of the aperture efficiency of the RIS, $\eta_{\rm ap}$, according to the following expression:
\begin{equation}
    \sigma_{\rm RIS} = \eta_{\rm ap}\frac{4\pi}{\lambda^2}D^4\cos(\theta_{\rm in})\cos(\theta_{\rm out}).
\end{equation}
The latter expression, which has been validated using full-wave simulations, implies that $D$ is the RIS variable to be determined, indicating its aperture size. The aperture efficiency will depend on many factors, including the phase quantization error and the unit cell insertion losses of the RIS. For passive structures, it is expected that a reflect-array can provide $50\%$ aperture efficiency. Additionally, the integration of the reconfigurable technology will degrade further the Radio Frequency (RF) performance of the metasurface. Based on previous works on reconfigurable technologies, at least $3$ dB of insertion losses are expected. Therefore, $25\%$ of aperture efficiency seems a reasonable limit to consider in the link budget calculations. We thus conclude that it would be required a $D=110$ mm RIS aperture size, corresponding to $10540$ unit elements. These numbers provide a more vivid picture on the scalability required for the RIS technology. 

\begin{figure}[t!]
	\centering
	\includegraphics[width=0.9\linewidth]{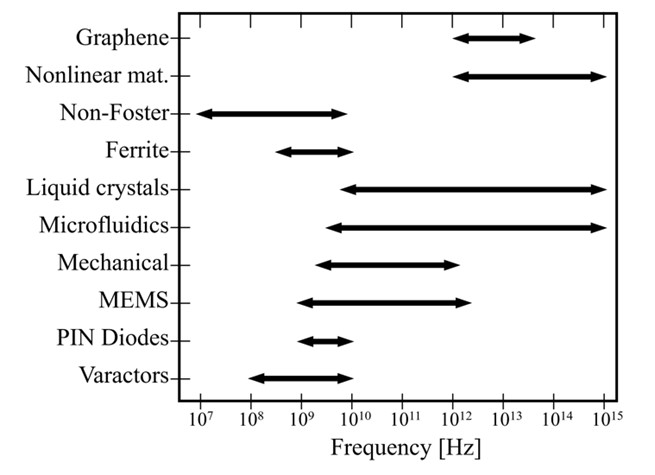}
	\caption{\small{Reconfigurable technologies for implementing RISs versus the operating frequency.}}
	\label{fig:RIS_vs_oper}
\end{figure}
\section{Challenges in THz RIS Design and Operation}
Beamforming with metasurfaces at millimeter waves, considering transmit-arrays, reflect-arrays, and phased arrays, has been widely studied (and still is) by the antenna community. Each technology has its own advantages and disadvantages, which translates into a balance among complexity, cost, power consumption and RF performance. In the context of 6G, two additional challenges arise: \textit{i}) new reconfigurable hardware components, capable of operating in the THz regime, need to be developed for integrating tunable-response RIS elements; \textit{ii}) Very large RIS sizes, with many thousands of elements, need to be assembled in a cost-effective way for overcoming the inherent free-space path loss at THz frequencies. 

\subsection{Hardware Design}
\begin{figure}[t!]
	\centering
	\includegraphics[width=0.95\linewidth]{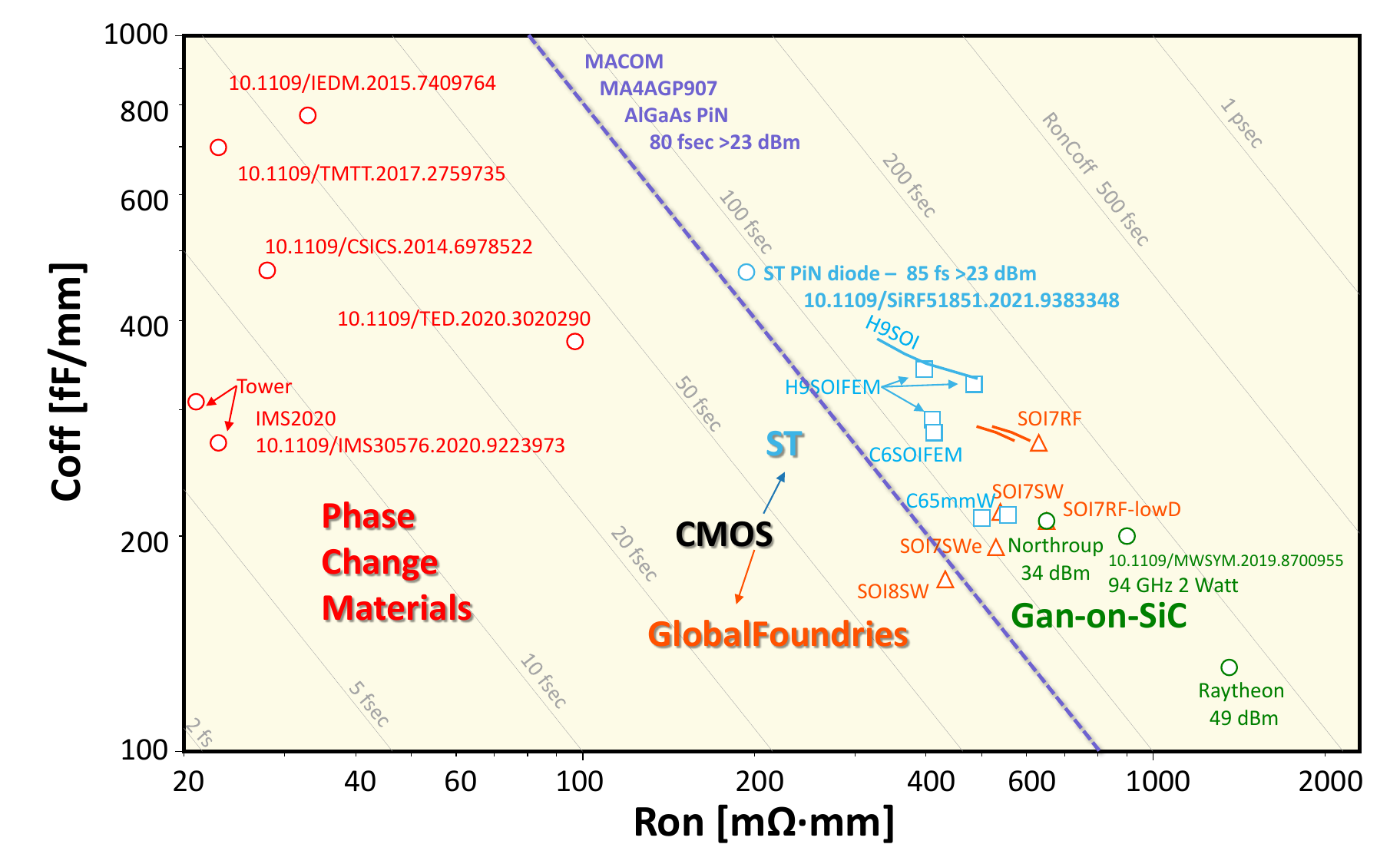}
	\caption{\small{$R_{\rm on}$ and $C_{\rm off}$ performance of reconfigurable technologies for implementing RISs (including publication DOIs) with respect to the maximum handling power and the switching time.}}
	\label{fig:Coff_Ron}
\end{figure}

As previously mentioned, there is a fundamental technological challenge related with finding a proper reconfigurable THz technology for realizing a cost-effective implementation of a switching mechanism. It is worth noting that the choice of the most suitable tuning technology to be adopted is highly dependent on the targeted frequency as well as the application. Switches based on Complementary Metal-Oxide Semiconductor (CMOS) constitute one promising technology, where the local bias circuitry can be realized within each RIS unit element. Therefore, each switch in each unit cell can be individually controlled, offering unparalleled flexibility in the configuration of the RIS. Moreover, CMOS-based switches lead to extremely low power consumption (in the order of tens of µW per unit cell), as opposed to diodes (several mW per unit cell). $45$ nm Radio-Frequency Silicon-On-Insulator (RF-SOI) CMOS technology is available on high resistivity substrates (resistivity of about few k$\Omega$×cm), which is expected to enable the integration, at affordable costs, of antenna elements (and unit cells) and low-loss CMOS-based switches with $R_{\rm on}$ (on resistance) and $C_{\rm off}$ (off capacitance) properties suitable for operation up to at least $300$ GHz. In~\cite{9731671}, a CMOS-based 98$\times$98 reflect-array at $265$ GHz was experimentally demonstrated. Additionally, Silicon-Germanium (SiGe) Bipolar CMOS (BiCMOS) technology, can be a simpler and cheaper alternative to RF-SOI to operate at least at $140$ GHz. Alternative emerging technologies, such as Phase Change Materials (PCM), are also gaining attention in the literature. In~\cite{Lin_coding}, an optically controlled metasurface integrating GeTe PCM was experimentally demonstrated around $300$ GHz. The currently available reconfigurable technologies for RIS design are summarized in Fig.~\ref{fig:RIS_vs_oper} and~\ref{fig:Coff_Ron} with respect to the operating frequency and the $R_{\rm on}$ and $C_{\rm off}$ performance. 

\begin{figure}[t!]
	\centering
	\includegraphics[width=0.95\linewidth]{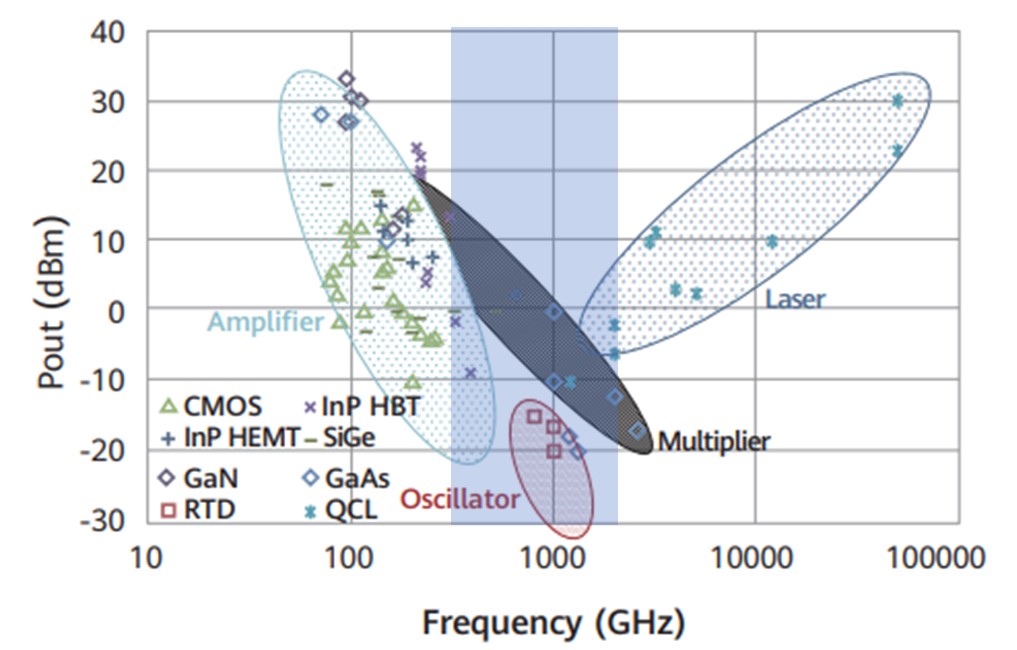}
	\caption{\small{Available transmit power levels by a THz source as a function of the operating frequency, considering different technologies for key components of the transmitter (THz gap)~\cite{THz_Huawei}.}}
	\label{fig:Thz_gap}
\end{figure}

\begin{figure}[t!]
	\centering
	\includegraphics[width=0.95\linewidth]{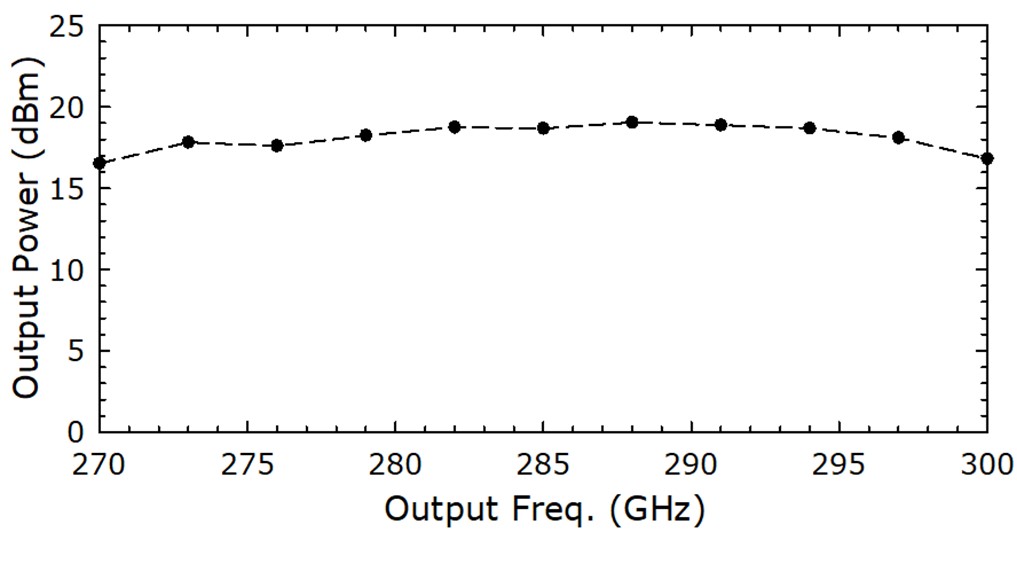}
	\caption{\small{Output transmit power at $285$ GHz using ACST's Schottky-based sub-THz power sources~\cite{9895924}.}}
	\label{fig:ACST}
\end{figure}

The other mentioned challenge of operating in the THz regime corresponds to the high path losses that limit the range of the wireless link, as shown in the link budget calculation in the previous Section~\ref{sec:THz_use_cases}. One approach to counteract this effect is to increase the source power, which turns out to be also quite challenging in the THz frequency band; see Fig.~\ref{fig:Thz_gap} including the achievable transmit power with various RIS technologies. Ongoing research is being carried out on utilizing high power sources with advanced modulation schemes in the THz regime. An example of a typical performance of a Schottky-based high power THz source is shown in Fig.~\ref{fig:ACST}. As depicted, the transmitting source, which is compatible with Amplitude-Shift Keying (ASK) modulation, is expected to have insertion losses of $3$ dB at the $285$ GHz frequency of operation. 

The radiation characteristics of the RIS constitute another bottleneck for the performance of an RIS-assisted wireless system~\cite{EURASIP_RIS}. In order to increase the aperture gain, the use of very large panels in terms of wavelengths is required, which results in additional stress to the energy consumption and cost requirements that will define the key performance indicators for the RIS hardware development. However, even assuming ideal switching mechanisms with low insertion loss and power consumption, the final aperture efficiency will depend on many fundamental RF design parameters, such as the number of quantization bits encoded in the aperture, the bandwidth where the required amplitude and phase reflection/transmission properties can be ensured (termed as the RIS bandwidth of influence in~\cite{EURASIP_RIS}), the maximum angular coverage provided by the aperture (indicating the RIS area of influence in~\cite{EURASIP_RIS}), and the beam stability with frequency (i.e., the beam squint). As an illustrative example, we consider the design frequency of $140$ GHz and an aperture taper of $10$ dB for an aperture diameter of $D = 80$ mm to produce a tilted plane wave at $\theta_{\rm out}=45^{\rm o}$. Using the PO-GO toolkit, it is possible to define an upper bound for the maximum gain of an idealized RIS aperture.
\begin{figure}[t!]
	\centering
        \includegraphics[width=0.95\linewidth]{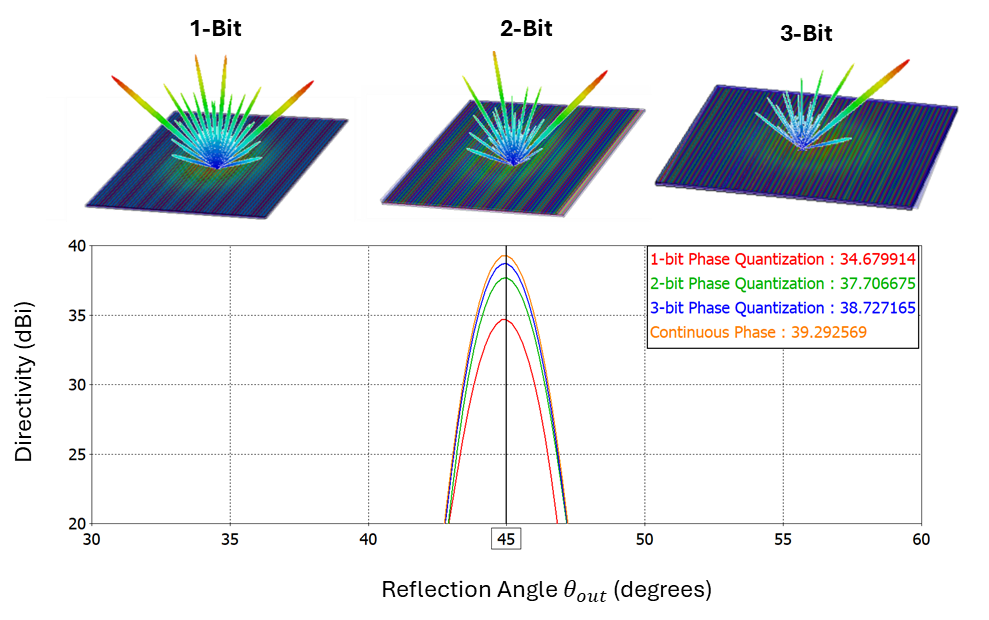}
	\caption{\small{Directivity of an RIS with $100\times100$ unit elements at the operating frequency of $140$ GHz as a function of the targeted reflection angle $\theta_{\rm out}=45^{\rm o}$ for different phase quantization levels.}}
	\label{fig:rad_pattern}
\end{figure}
\begin{figure}[t!]
	\centering
        \includegraphics[width=0.95\linewidth]{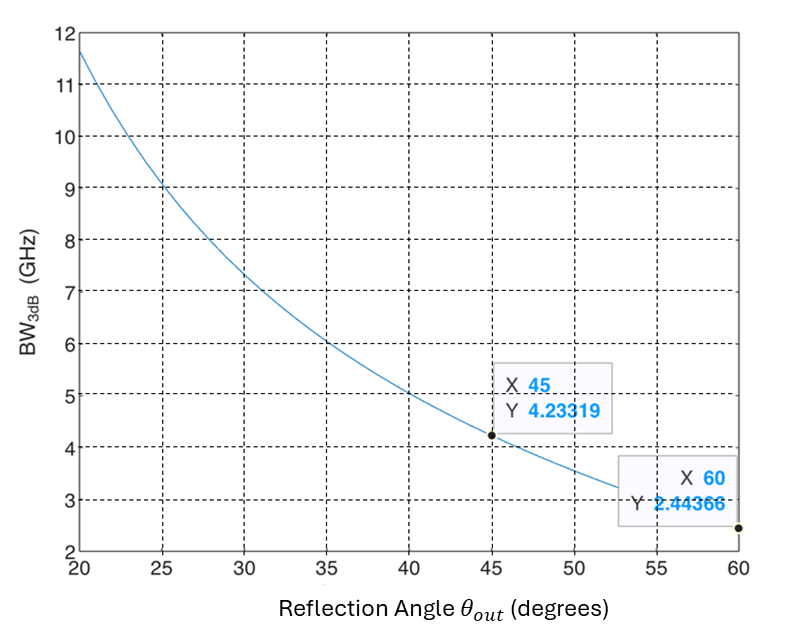}
	\caption{\small{The $3$ dB bandwidth, ${\rm BW}_{\rm 3dB}$, for beam squint at the operating frequency of $140$ GHz as a function of the reflection angle $\theta_{\rm out}$.}}
	\label{fig:3db_bandwidth}
\end{figure}
Figure~\ref{fig:rad_pattern} depicts the idealized gain of this aperture considering different quantization bits for the RIS element phase response. It is worth noting that the $2$-bit resolution provides a good trade-off between RF performance and integration complexity of multiple switches into a single RIS unit cell. Another important aspect for the metasurface design, which is many times overlooked in high-level analyses, is the fact that the intrinsic phase wrapping used in the RIS design implies the occurrence of the beam squint, which, for very narrow beams, can have significant impact on the system performance. In fact, one can quantify the $3$ dB bandwidth, ${\rm BW}_{\rm 3dB}$, due to the angle shift caused by the beam-squint effect as a function of the reflection angle $\theta_{\rm out}$, as shown in Fig.~\ref{fig:3db_bandwidth}. The parameter ${\rm BW}_{\rm 3dB}$ corresponds to the operation bandwidth where the gain is maintained within $3$ dB below the central frequency gain. It can be observed from this figure that, this bandwidth's value for beam squint is $4.2$ GHz ($3\%$) at $\theta_{\rm out}=45^{\rm o}$ and $2.4$ GHz at an angle of $\theta_{\rm out}=60^{\rm o}$ ($1.7\%$), showcasing the importance of further investigating techniques to overcome the impairments induced by the beam-squint effect.

\subsection{Signal Processing}
The programmable reflective beamforming capability of RISs necessitates the knowledge of parameters related to the channels~\cite{Tsinghua_RIS_Tutorial} between an RIS and the end nodes that wish to profit from its operation. For example, in Fig.~\ref{fig:NLOS_RIS}, the entity tasked to dynamically optimize the RIS beamforming needs to frequently estimate the wireless channels between the RIS and the BS as well as the RIS and the mobile terminal. This estimation needs to take place within the channel coherence time with as less timing overhead and complexity as possible to allow enough time for communications. Recall from the link-budget analysis in Section~\ref{sec:link_budget} that, in THz frequencies, the RIS may include thousands of unit cells, resulting in extremely large channel matrices, whose timely estimation is challenging. Therefore, efficient estimation algorithms for large channel matrices and low latency RIS control protocols need to be developed, which need to be realized with hierarchical RIS beam codebooks for both near- or far-field communications. In addition, efficient schemes for RIS-based THz localization~\cite{Keykhosravi2022infeasible} and integrated sensing communications~\cite{RIS_ISAC_SPM} are necessary.

Multi-functional RISs, with some of them including active components for reflection amplification and/or impinging signal sensing~\cite{RISoverview2023}, or even for transmission/reception~\cite{Shlezinger2021Dynamic}, are lately receiving considerable attention as candidate solutions for extending signal coverage and/or efficiently dealing with channel estimation. Their potential for THz wireless systems needs to be however justified via efficient and scalable hardware designs as well as adequate signal processing algorithms.

\section{Conclusion}
In this paper, we overviewed some of the most significant hardware design and signal processing challenges associated with THz RISs, as they have been also identified by the ongoing TERRAMETA project. We presented a preliminary analysis of the impact of RISs on the overall THz link budget and system performance. We found that an aperture size of $D=110$ mm would be required for an RIS-enabled $100$ m wireless link at $140$ GHz, which would amount to $10540$ elements. These numbers already provide a rough picture on the scalability of the RIS technology towards higher frequency ranges. We also discussed challenges for RIS-based beamforming, including output power, reconfiguration technology, phase quantization, and beam squint effects.

\section*{Acknowledgments}
This work was supported by the the Smart Networks and Services Joint Undertaking (SNS JU) under the European Union's Horizon Europe research and innovation programme under Grant Agreement No 101097101, including top-up funding by UK Research and Innovation (UKRI) under the UK government’s Horizon Europe funding guarantee.

\bibliographystyle{IEEEtran}
\bibliography{references}
\end{document}